\documentclass[journal]{IEEEtran}
\usepackage{cite}

\ifCLASSINFOpdf
  \usepackage[pdftex]{graphicx}
  \graphicspath{{plots/}}
\else
\fi

\usepackage{float}

\usepackage[hidelinks]{hyperref}

\usepackage{subcaption}
\usepackage[most]{tcolorbox}
\usepackage{booktabs}

\begin{document}
%
\title{How international are international computing conferences? \\ --- An exploration with systems research conferences}

%
%
%


\author{Pedro~Garcia~Lopez, 
    Marina~López~Alet, 
    Usama~Benabdelkrim~Zakan, 
    and~Anwitaman~Datta
\thanks{P. García, M. López and U. Benabdelkrim are with Universitat Rovira i Virgili, Spain.}%
\thanks{A. Datta is with Nanyang Technological University, Singapore.}%
}

%
%

\markboth{}%
{Shell \MakeLowercase{\textit{et al.}}: Bare Demo of IEEEtran.cls for IEEE Journals}
%



\maketitle

\begin{abstract}
    In recent years, Asia's rapid growth in research output has been reshaping the computing research landscape. What was once a two-block system (America and Europe) is evolving into a multipolar world with three major hubs: America, Europe, and Asia. To study these pivotal changes and evaluate international diversity, we have analyzed the past 13 years of 13 international systems research conferences: ASPLOS, NSDI, OSDI, SIGCOMM, ATC, EuroSys, ICDCS, Middleware, SoCC, CCGRID, IC2E, IEEE Cloud and EuroPar. Our analysis focuses on accepted papers and  participation in the Program Committee, grouping the results by region (America, Europe, and Asia). Surprisingly, we find a pronounced historical imbalance in international diversity among top-tier systems conferences (ASPLOS, OSDI, NSDI, SIGCOMM). While most other conferences have progressively reflected Asia’s growing research presence over the past decade, this group has shown a noticeable adjustment only in the recent four years. We also identify persistent rigidities in how program committee (PC) diversity adapts to shifts in accepted paper origins, with a consistent under-representation of researchers from Asian organizations in many PCs.

\end{abstract}

%
\IEEEpeerreviewmaketitle

\section{Introduction}

The enormous economic growth of Asia over the past few decades~\cite{Lee2010} is expected to drastically change the power equilibrium in the XXI century. Previous prospects~\cite{Lee2010} forecast that Asia will account for two-thirds of the world economy by 2030, almost double the share of 34\% from 2009. This rapid economic growth is reshaping the scientific landscape in tandem, resulting in a rising number of scientific publications from Asia. Previous bibliometric studies~\cite{Jang2014} have already analyzed this surge in publication and citation growth from Asia, placing emphasis on their primary focus on technology, as well as applied science.


Within the field of computer science, bibliometric studies on geographical diversity~\cite{jiao2021bibliometric} reveal that the United States remains the leading contributor, while China and other Asian nations have experienced sustained growth over the past decade, positioning the region as a major player in global research. However, most prior investigations have focused primarily on journal publications, despite a popular perception, as encapsulated in Mattern’s observation~\cite{mattern2008bibliometric} that conferences play an exceptional and often dominant role in the dissemination of knowledge within computer science.



In this paper, we analyze how the computer systems research community is responding to the profound scientific and geopolitical shifts emerging from Asia. We examine international scientific diversity, understood as the participation of researchers from different countries and continents in both accepted papers and program committees. Our study focuses on computer systems research, a field of increasing geopolitical relevance driven by the global expansion of cloud computing and artificial intelligence. To this end, we analyze thirteen major international conferences in this area over the period 2012–2024: ASPLOS, NSDI, SIGCOMM, OSDI, ATC, EuroSys, ICDCS, Middleware, SoCC, CCGRID, IC2E, IEEE Cloud, and EuroPar.

They represent popular conferences in the field that have been active in the past (at least) thirteen years. According to the CORE ranking~\cite{CORE}, we selected conferences ranked as A*, A, and B. ASPLOS, OSDI and SIGCOMM are ranked A*, and EuroSys, ATC, ICDCS, Middleware, CCGRID and NSDI received A rankings during the first eleven years, while IEEE Cloud and EuroPar were ranked as B. We also included unranked yet popular conferences, namely IC2E and SoCC. 

To carry out our analysis, we created a crawler~\cite{crawler} that retrieves information from conference websites and three major research resources: DBLP~\cite{DBLP}, Semantic Scholar~\cite{SemanticScholar}, and OpenAlex~\cite{OpenAlex}. Using the collected data, we investigate how the systems research community is adapting to ongoing geoeconomic shifts and, in particular, how these transformations are reflected in the growing internationalization of research participation. We have developed a public website~\cite{WebDataVisualizer} which facilitate visual exploration of the information; moreover all the underlying datasets are openly available on Dataport~\cite{DataPort}, and the complete data analysis pipeline is accessible through CodeOcean~\cite{CodeOcean}.

As recently observed in~\cite{Economist2024}, the rapid rise of Chinese and Asian science should lead to a more diverse scientific community, even if it disrupts existing balances of power distribution. Our exploration reveals interesting insights related to international diversity in accepted papers, and Program Committees across the studied conferences. Although some conferences such as ICDCS, ATC, EuroSys, IEEE Cloud and CCGRID have adapted to Asian growth in the last decade, we identified a group of  top-tier conferences (ASPLOS, NSDI, SIGCOMM, OSDI)  that have only started to adapt in the recent four years.


\section{The crawler and the data}
Our crawler gathers information from three main sources. DBLP provides structured metadata that allows us to identify the papers published each year for the selected conferences, including titles, authors, and publication years. OpenAlex, on the other hand, supplies information about author affiliations and institutions at the time of publication. Finally, Semantic Scholar allows us to extract additional information from papers, such as abstracts or TLDRs.
It is important to note that the crawler does not automatically extract data about program committee (PC) members due to the heterogeneity of conference websites. The structure and location of PC member lists often vary between years and venues—sometimes being split across multiple subpages or entirely missing for certain editions—which makes automated extraction unreliable. Therefore, all PC member data were collected manually to ensure completeness and accuracy.

In total, the dataset includes 9,712 accepted papers (from all the conferences under study) and 14,996 non-deduplicated program committee members, corresponding to 6,917 unique individuals. While the information on papers and their authors was automatically collected by the crawler, data on program committee members were manually compiled due to the lack of a consistent structure across conference websites.

We mapped both papers and program committee members to continents based on the affiliations of their authors or members. 
For papers, this process was performed automatically by the crawler at the time of data collection, while for program committee members it was carried out manually using the same assignment rules. 
The affiliation assignment followed the following heuristic: if most authors belong to the same continent, the paper (or member) is assigned accordingly; in cases of ties (equal representation from multiple continents), the assignment is made uniformly at random. 
If most authors lack institutional or country information, the assignment is based on the majority continent among the remaining authors. 
Finally, if no country information is available for any of the authors, the entity remains unassigned.

Some mappings had to be performed manually for papers and program committee members whose affiliation information was missing from the APIs used (DBLP and OpenAlex). 
In these cases, the missing data were retrieved directly from the official websites of the corresponding conferences for each year. 
This approach ensured that the information was obtained from authoritative primary sources, guaranteeing the accuracy and completeness of the affiliations. 
Although a small percentage of records still lacked affiliation data, this limitation does not significantly affect the overall geographical distribution observed in the study. 
Therefore, our subsequent analyses rely on the assumption that this small fraction of missing data does not introduce a systematic bias in the observed trends.

Although the data obtained from the APIs were occasionally incomplete or contained minor inconsistencies, these cases were corrected whenever possible using information retrieved manually from the official conference websites. 
Due to these design decisions, approximately 5\% of the papers in each conference were assigned to a continent despite most of their authors lacking affiliation data. 
Similarly, no conference had more than 5\% of papers associated with multiple continents, indicating that over 90\% of the papers analyzed were assigned unambiguously to a single continent. 

Although the dataset may still contain minor imperfections that could slightly affect quantitative results, we have manually vetted its consistency using multiple randomized samples to minimize such effects. 
Therefore, to the best of our assessment, any residual inaccuracies are unlikely to have a material impact on the qualitative conclusions drawn from this study.

\section{The increasing participation of Asia}

\begin{figure*}[ht]
    \centering
    \includegraphics[width=\textwidth]{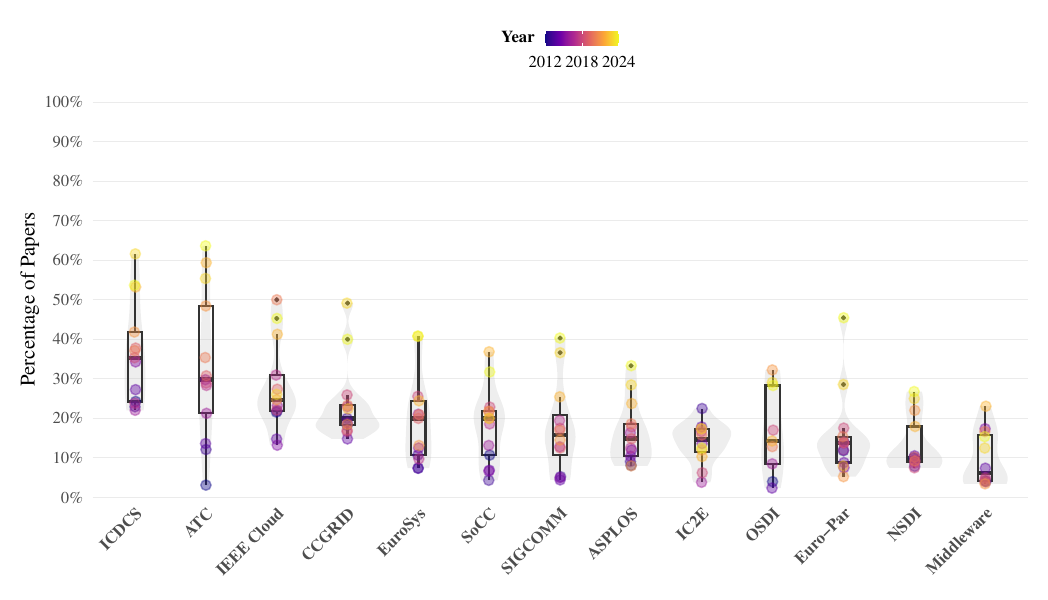}
    \caption{Distribution of Asian publications.}
    \label{fig:asian_trend}
\end{figure*}


Figure~\ref{fig:asian_trend} illustrates the variability in the proportion of Asian papers presented at various conferences over time. While the proportion of Asian papers has remained consistent at certain conferences, such as IC2E and Middleware, most conferences show increasing Asian participation during the past decade. In particular, we can see how ICDCS, ATC, IEEE Cloud, CCGRID and EuroSys show a growing trend that it is peaking in the last years.


This increase in the proportion of Asian papers suggests a proportional (though not necessarily absolute) decline in European and North American publications at most conferences. 
It is noteworthy that a group of top-tier conferences historically dominated by American papers (NSDI, ASPLOS, OSDI, SIGCOMM, SoCC) has only begun to reflect the growth of Asian research within the last four years. In these venues, the share of North American contributions has declined in recent years from roughly 80\% to about 55\% for NSDI and ASPLOS, and to 52\% and 48\% for OSDI and SIGCOMM, respectively.

Another noteworthy case is USENIX ATC, a top-tier conference where Asian participation—primarily from China—rose from nearly 5\% in 2012 to about 63\% of accepted papers in 2024. The last year US only accounted for 27\% of the papers while it was 78\% in 2012. For ATC, shift had been steady over the last decade, in contrast to the last four years for some of the other top conferences. 

As seen in Figure~\ref{fig:asian_trend},  the rest of the conferences have accommodated Asian growth during the last decade, with less pronounced changes than the top conferences.

\section{International diversity}
In our analysis, the international diversity of a conference is assessed by examining the geographical distribution (at the granularity of continents) of the institutions affiliated with the authors of papers published in different years. For each conference, we calculate the percentage of papers attributed to each continent. We consider international diversity to be low when a single continent dominates with a high percentage of papers, with little representation from other continents. Conversely, a conference is considered more internationally diverse when the percentages of papers across continents are more balanced, indicating a more equitable geographical distribution among the affiliations of the authors.

\begin{figure*}[t]
  \centering
  \includegraphics[width=\textwidth]{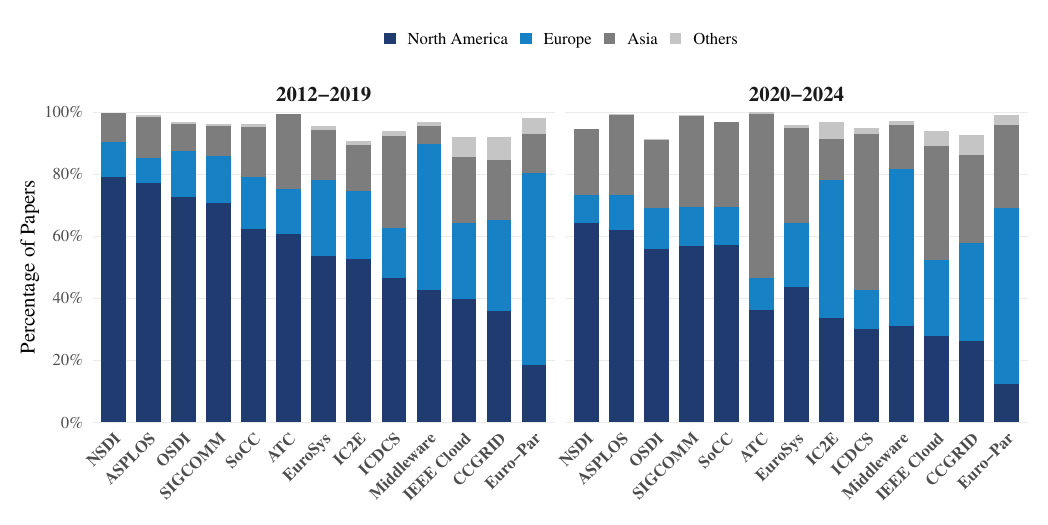}
  \caption{Continent diversity in accepted papers divided into two ranges of years (2012-2019) and (2020-2024).}
  \label{fig:accepted_papers_distribution}
\end{figure*}

Figure~\ref{fig:accepted_papers_distribution} shows how papers from different continents have been distributed across the various conferences. We have decided to add the data in two different ranges of years. We have chosen these ranges based on the general evolution of the data for the conferences. Analysis of the data reveals a recent surge in Asian publications in the last four years, as shown in Figure~\ref{fig:asian_trend}. The first range of years corresponds to the aggregated data for 2012-2019, while the second range (2020-2024) reflects the most recent years. 

Figure~\ref{fig:accepted_papers_distribution} clearly illustrates the substantial changes that have taken place within a relatively short period of time. This information indicates that the majority of conferences have been adapting to current changes in the field of computer science research. Previously underrepresented conferences, such as NSDI, where the majority of papers came from North America and Asian papers represented approximately 10\% of all papers published between 2012 and 2019, have now increased their diversity. Over the last four years, Asian papers have represented approximately 20\% of the papers published. 

This shift in focus is not exclusive to NSDI; other conferences such as ASPLOS, OSDI, SIGCOMM, and SoCC have also undergone similar transitions toward greater diversity. 


\begin{figure*}[h]
  \centering
  \includegraphics[width=\linewidth]{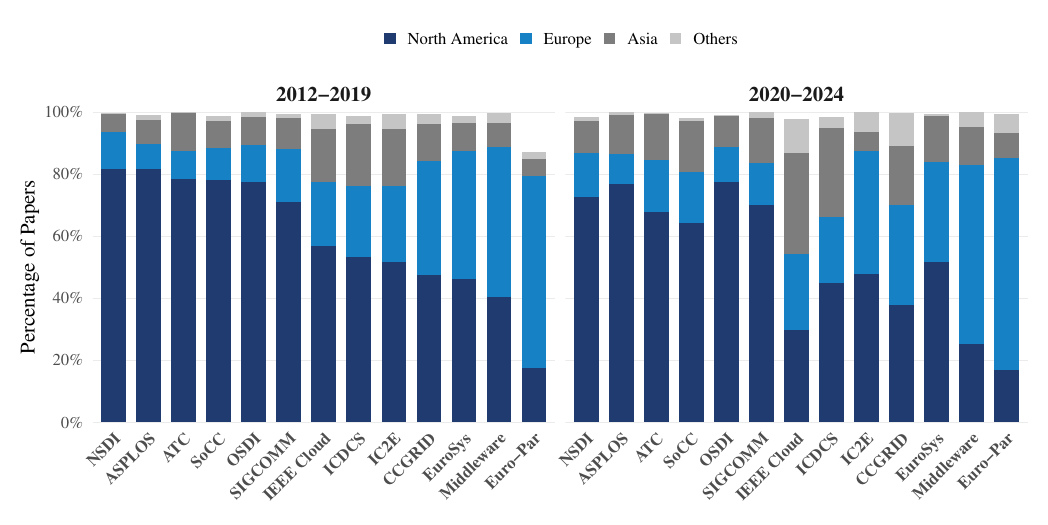}
  \caption{Continent diversity in Committee Members divided into two ranges of years (2012-2019) and (2020-2024).}
  \label{fig:committee_distribution}
\end{figure*}

Figure~\ref{fig:committee_distribution} illustrates the international distribution of the program committee across all conferences, aligning with the same range of years as Figure~\ref{fig:accepted_papers_distribution}. As can be seen, in the first range of years (2012-2019), conferences such as NSDI, ASPLOS, ATC, SoCC, and OSDI are predominantly North American, with Europe and Asia representing only between 10\% and 20\% of the program committee members. In contrast to this, we can see how conferences such as Middleware and Euro-Par had a large European representation. 

However, when we examine the subsequent period from 2020 to 2024 for program committee composition, we observe a lack of substantial change, in contrast to what was observed for accepted papers in Figure~\ref{fig:accepted_papers_distribution}. It is apparent that Asian researchers' participation in program committees is not yet in parity and is underrepresented with respect to their contributions in terms of publications (Figure~\ref{fig:asian_trend}). This is evident in ATC, which features a significant proportion of Asian publications, yet reflects only a small percentage of Asian members in the committee. This will be further analyzed in section~\ref{sec:PC vs. Accepted Papers}.


To quantify the diversity of conferences, we used the Gini-Simpson index~\cite{SimpsonIndex}, a variation of the Simpson diversity index widely used in biology to measure species biodiversity in a given area. We applied it to determine the diversity of the conferences across three categories (North America, Europe, Asia) based on the number of papers published by each continent in the different conferences. 

\begin{equation}
  1 - D = 1 - \sum_{i=1}^{R} p_i^2
  \label{eq:gini-simpson}
\end{equation}

The Gini-Simpson index is calculated using the formula shown in Equation~\eqref{eq:gini-simpson}, where \( R \) represents the number of available categories, in this case, three (North America, Europe, and Asia), and \( p_i \) represents the proportion of papers from each continent relative to the total number of papers for each conference (excluding papers not assigned to any continent).
This index indicates the probability that two papers randomly selected from the same conference belong to different continents. Thus, a higher index value indicates greater diversity, meaning that the articles are more evenly distributed across continents, while a lower value suggests lower diversity, implying dominance by one or two continents.


\begin{table}[h]
\centering
\begin{tabular}{lcccc}
\hline
Conference & \multicolumn{2}{c}{Cumulative} & \multicolumn{2}{c}{Year 2024} \\
\cmidrule(lr){2-3} \cmidrule(lr){4-5}
 & Papers & PC & Papers & PC \\
\hline
NSDI & 0.405 & 0.319 & 0.527 & 0.564 \\
ASPLOS & 0.467 & 0.346 & 0.555 & 0.419 \\
SIGCOMM & 0.484 & 0.431 & 0.594 & 0.509 \\
EuroPar & 0.491 & 0.401 & 0.595 & 0.497 \\
OSDI & 0.510 & 0.364 & 0.577 & 0.454 \\
SoCC & 0.522 & 0.416 & 0.589 & 0.582 \\
Middleware & 0.574 & 0.563 & 0.592 & 0.526 \\
ATC & 0.595 & 0.439 & 0.502 & 0.390 \\
IC2E & 0.598 & 0.590 & 0.498 & 0.512 \\
EuroSys & 0.610 & 0.599 & 0.582 & 0.532 \\
ICDCS & 0.621 & 0.605 & 0.570 & 0.642 \\
IEEE Cloud & 0.654 & 0.626 & 0.587 & 0.656 \\
CCGRID & 0.658 & 0.614 & 0.647 & 0.631 \\
\hline
\end{tabular}
\caption{Gini-Simpson’s Diversity Index for Accepted Papers and Committee Members. Both the index calculated for the thirteen years combined and for the last year studied (2024) are shown.}
\label{tab:simpson_diversity}
\end{table}


Table~\ref{tab:simpson_diversity} shows the diversity results obtained by applying the Gini-Simpson index, both in the overall data (aggregated from 2012 to 2024) and the diversity indices for the last year (2024). The conferences in the table are sorted from those with the least diversity in the accepted papers of the aggregated global data to those with the greatest diversity. When observing the aggregated diversity results, it is clear that conferences such as CCGRID (0.658), IEEE Cloud (0.654), and ICDCS (0.621) have greater diversity in accepted papers compared to conferences such as NSDI (0.405), ASPLOS (0.467), and SIGCOMM (0.484). This indicates that ICDCS, CCGRID and IEEE Cloud have a more balanced geographic distribution of accepted papers, suggesting a more international scope compared to NSDI, ASPLOS, and SIGCOMM. 

However, when examining last year's data, it becomes evident that the publication diversity indices are more closely aligned. NSDI (0.527), ASPLOS (0.555), and SIGCOMM (0.594) demonstrate more closely aligned results, with SIGCOMM exhibiting a higher score compared even to some of the historically most diverse conferences (based on the aggregated data), including ICDCS (0.570) and IEEE Cloud (0.587) in 2024. 


The Simpson index values obtained for all the years aggregated for the program committee appear to follow a similar pattern to those obtained for the accepted papers. Conferences such as CCGRID (0.614), IEEE Cloud (0.626), and ICDCS (0.605) have more diverse committees compared to NSDI (0.319), ASPLOS (0.346), and OSDI (0.364). A similar trend is evident when examining these indices over the past year (2024), as they have become more closely aligned, indicating a concerted effort by the Program Committees to align with the distribution represented by the accepted papers. ATC (0.390) has achieved the least diversity in its program committee, as the committee has not been able to adapt to the large number of Asian papers that have been accepted this past year. This gap between the papers and the committee is analyzed in greater depth in section~\ref{sec:PC vs. Accepted Papers}.


As observed, some conferences are clearly more diverse than others. One might hypothesize that the location where a conference is held might influence the proportion of papers and members from a particular continent. Although a more thorough investigation of the causes of disparity remains an open issue, some preliminary and informal analysis of the various conferences showed that this is not entirely the case.

What is true is that conferences such as NSDI, ATC, OSDI and SoCC have only ever been held in North America, which could explain their strong American presence. However, conferences such as ASPLOS, EuroSys and SIGCOMM challenge this argument. EuroSys is always held in Europe, yet it has a high proportion of American participants. However, SIGCOMM and ASPLOS also move to other continents, yet remain predominantly American in representation. 

Further investigation reveal that a significant percentage of Asian and European papers presented at less diverse conferences such as NSDI, ASPLOS, and SIGCOMM have at least one author affiliated with a North American institution or company. Approximately, more than 30\% of the European papers presented at these conferences also had a North American author. As for Asian papers, at conferences such as NSDI, SIGCOMM, and EuroSys, more than 40\% of published Asian papers had an author affiliated with a North American institution or company.

\section{Geographical Disparities Between Program Committees and Accepted Papers}
\label{sec:PC vs. Accepted Papers}

This section analyzes the \emph{geographical representation gap} between Program Committee (PC) members and accepted papers across major conferences from 2020 to 2024. For each venue and continent, the gap is computed in percentage points (pp) as:

\[
\text{Gap} = \text{Committee}_{\%} - \text{Papers}_{\%}.
\]

A positive value indicates over-representation in the PC relative to accepted papers, while a negative value denotes under-representation.

\begin{figure*}[t]
    \centering
    \includegraphics[width=\textwidth]{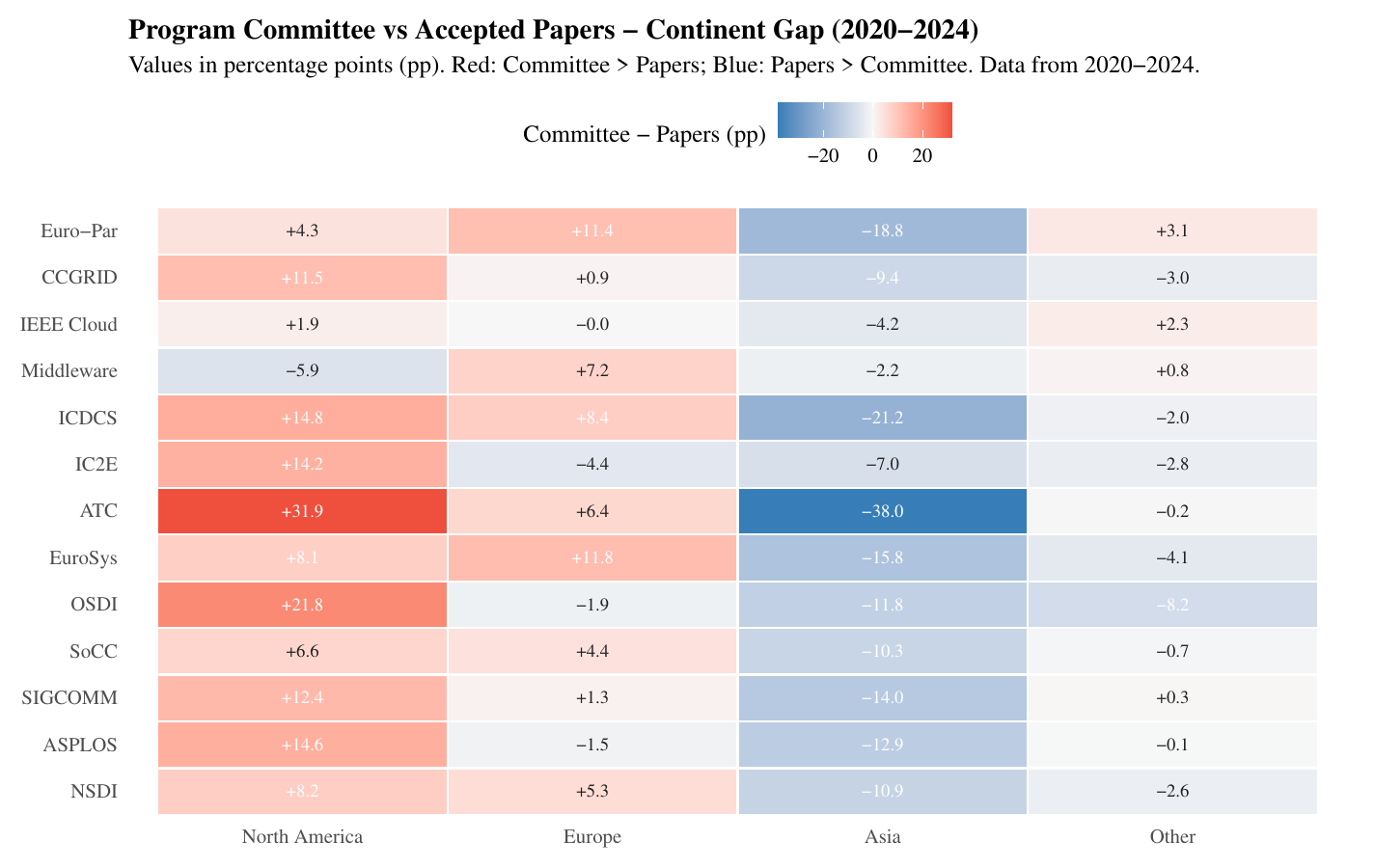}
    \caption{Program Committee vs.\ Accepted Papers—Continent Gap (2020–2024). Values are in percentage points (pp). Red indicates higher PC representation relative to accepted papers, blue indicates the opposite.}
    \label{fig:papers_vs_committee_continent_gap}
\end{figure*}

To compute these gaps, the data were first grouped by \textit{Conference} and then by \textit{Continent} to count the number of records for each combination. These counts were then converted into percentages within each conference, both for PC members and accepted-paper authors. Finally, the geographical gap was obtained as the difference between both percentages (\(\text{Gap} = \text{Committee}_{\%} - \text{Papers}_{\%}\)). The resulting heatmap (Figure~\ref{fig:papers_vs_committee_continent_gap}) visualizes these gaps by conference and continent, with titles and subtitles indicating the range of years considered (2020–2024).

Figure~\ref{fig:papers_vs_committee_continent_gap} reveals persistent geographical disparities across the 2020–2024 period. North America remains consistently over-represented in Program Committees across nearly all venues. The strongest positive gaps appear in ATC (+31.9\,pp), OSDI (+21.8\,pp), ASPLOS (+14.6\,pp), ICDCS (+14.8\,pp), IC2E (+14.2\,pp), SIGCOMM (+12.4\,pp), CCGRID (+11.5\,pp), and EuroSys (+8.1\,pp). Only Middleware (–5.9\,pp) shows a slight under-representation, standing as an outlier among the otherwise dominant North American presence.


In contrast, Asia is systematically under-represented across almost all conferences. The most pronounced gaps occur in ATC (–38.0\,pp), ICDCS (–21.2\,pp), Euro-Par (–18.8\,pp), EuroSys (–15.8\,pp), and SIGCOMM (–14.0\,pp). Other venues such as OSDI (–11.8\,pp), SoCC (–10.3\,pp), ASPLOS (–12.9\,pp), and NSDI (–10.9\,pp) also show consistent negative differences, confirming that Asian institutions contribute significantly fewer PC members relative to their share of accepted papers. Middleware (–2.2\,pp) remains the only venue where the under-representation is minor. 



Overall, the 2020–2024 data confirm a persistent asymmetry: Program Committees are predominantly composed of members from North America, while Asia continues to be significantly under-represented across virtually all major conferences. This imbalance highlights the enduring concentration of decision-making and peer-review power in North American institutions, while Asian researchers remain comparatively under-represented despite their increasing publication output.



\section{Tech Companies in academic research}

\begin{figure*}[t]
    \centering
    \includegraphics[width=\textwidth]{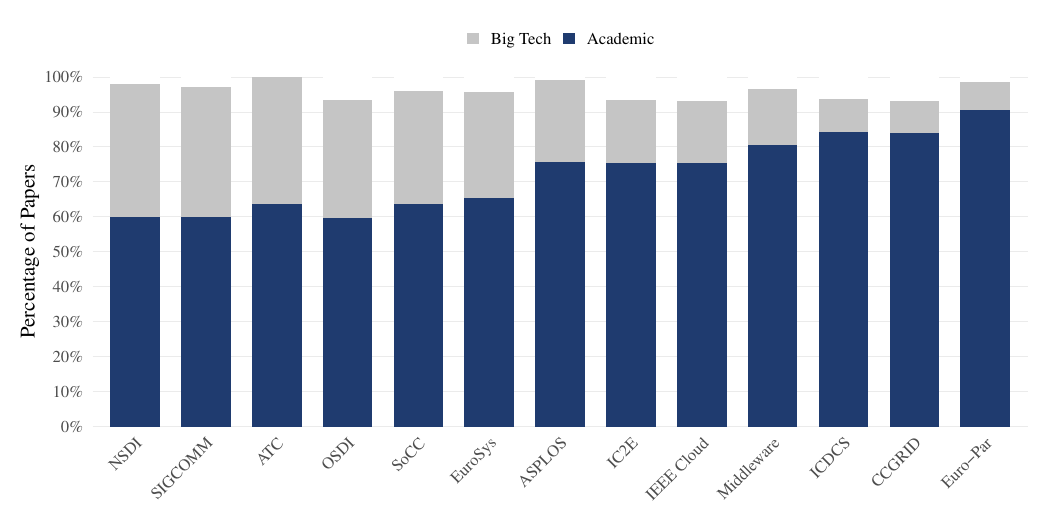}
    \caption{Papers with authors belonging to a big tech company vs papers with authors exclusively academic.}
    \label{fig:tech_companies_papers}
\end{figure*}

As illustrated in Figure~\ref{fig:tech_companies_papers}, the proportion of papers published at each conference with at least one author affiliated with a major technology company is shown, in contrast to papers authored solely by members of academic institutions. The selected companies represent the leading technology companies actively engaged in research. These include IBM, Microsoft, Google, Amazon, Facebook, Apple, Intel, Oracle, Cisco, HP, NVIDIA, Meta, VMware, Netflix, Uber, AMD, Huawei, Alibaba, ByteDance, Samsung, Xiaomi, ARM, and SAP, among others. The figure illustrates the tendency of conferences such as NSDI, SIGCOMM, ATC, and OSDI to be the preferred venues for large technology companies to publish, whereas conferences like IEEE Cloud, CCGRID, and Euro-Par predominantly feature papers authored exclusively by members of academic institutions.

When analyzing Figure~\ref{fig:tech_companies_papers} in conjunction with Figure~\ref{fig:accepted_papers_distribution}, it is observed that conferences with a higher percentage of papers from U.S.-based authors also tend to have a greater share of publications from major technology companies (e.g., NSDI, SIGCOMM, OSDI). Conversely, conferences with a more pronounced international character tend to publish papers primarily from academic institutions (e.g., ICDCS, IEEE Cloud, and CCGRID).

Our analysis shows that program committee members are predominantly affiliated with academic institutions, representing more than 75\% of participants in most conferences. Among them, ICDCS, CCGRID, and Euro-Par exhibit the lowest levels of participation from large technology companies, whereas SoCC, ATC, OSDI, and NSDI display the highest corporate representation.

\section{Conclusions}

We analyzed data from thirteen leading conferences in computer systems research over the past thirteen years to assess their levels of international diversity. To the best of our knowledge, this is the first work that studies the geographical diversity of computer science conferences regarding accepted papers and program committees.

Previous bibliometric studies—mostly focused on journals—have highlighted the continued dominance of the United States in research, alongside a growing contribution from China and other parts of Asia. Our findings confirm the sustained leadership of the United States and the increasing participation of Asian institutions. However, we also reveal three notable results: (i) large disparities in diversity adaptation among conferences, (ii) significant involvement of major technology corporations in the less diverse venues, and (iii) clear under-representation of Asian researchers in PCs. 

First of all, a group of top-tier conferences including NSDI, OSDI, ASPLOS, and SIGCOMM have traditionally shown lower diversity, with American participation bordering 80\% in the past decade. These conferences only started to accommodate a significant number of Asian contributions in the last three or four years. Those recent changes are profound, reducing American participation to around 50\% at the expense of Asian growth in just four years. In stark contrast, the inclusion of Asian research papers have been more gradual in the rest of conferences during the past decade.

We note that this imbalance is coincident with dominant presence of technology giants in the less diverse research conferences. In those conferences, around 40\% of papers come from Big Tech companies. This creates intense competition, making it increasingly difficult for academic researchers worldwide to participate on equal footing. While our current study does not confirm a causation, it generates evidence to hypothesize that the relative lack of international diversity in the last decade in these conferences is driven by a combination of two factors: (i) their very high reputation in the research community makes them the preferred choice for industry-based researchers to participate in, and that in turn (ii) makes these venues more competitive and thus harder for academic contributors to participate in. The rise of Big Tech in China could in that case also explain the sudden rise of Asian papers in some of the top conferences in the recent years. We reiterate that our current analysis does not confirm these causalities, but that the current observations lead us to hypothesize this as a potential factor which merits further study.

Finally, we have reported certain rigidities in most conferences to adapt their PCs to the new research landscape. In general, we observe a clear under-representation of Asian researchers in most PCs (according to their participation in papers). 

USENIX ATC is perhaps an extreme case to understand what is happening in the field. ATC evolved from  2\% of Asian papers in 2012 to 63\% in 2024. Yet, in 2024, 75\% of the PC was American and only 17.6\% was Asian. In 2024, 20.78\% of papers came from Asian Big Tech companies and 15.58\% from American Big Tech companies. Interestingly, the organizers announced the conclusion of the event in July~2025, following the final edition of USENIX ATC'25~\cite{usenix_atc_announcement}. 

The growing importance of technological innovation in AI and cloud research and fierce geopolitical competition in these fields anticipate greater international diversity in the next years. We hope that the presentation of our analysis encourages reflection within the community toward maintaining geographical diversity and fostering more equitable global participation in science, while continuing to uphold—and even enhance—the quality of the research presented in these venues.

\ifCLASSOPTIONcaptionsoff
  \newpage
\fi



%

\bibliographystyle{IEEEtran}
\bibliography{references.bib}

%








\end{document}